\documentstyle[12pt]{article}
\def\bl{\Biggl\{}
\def\br{\Biggr\}}
\def\bpl{\Biggl(}
\def\bpr{\Biggr)}

\def\d{\delta}
\def\D{\Delta}

\def\M{{\cal M}}

\def\gbig {\hbox{\Large\it g}}

\def\l{\lambda}

\def\so{\emptyset}

\def\z{\zeta}

\def\s{\sigma}

\def\z{\zeta}

\def\vphi{\varphi}
\def\w{\omega}

\def\hf{\frac{1}{2}}
\def\der{\partial}

\def\bq{\begin{equation}}
\def\eq{\end{equation}}
\def\brr{\begin{eqnarray}}
\def\err{\end{eqnarray}}
\def\ba{\left(\begin{array}}
\def\ea{\end{array}\right)}

\def\ba{\left(\begin{array}}
\def\ea{\end{array}\right)}
\begin{document}
\begin{flushright}
CERN-TH/95-289\\
\end{flushright}

\vspace*{3.5cm}
\begin{center}
{{\large\bf Towards Evaluation of Stringy Non-Perturbative Effects}}

\vspace*{6.0ex}
{\large Ram Brustein${}^{a,b,*)}$ and Burt A. Ovrut${}^{c)}$}

\vspace*{1.5ex}
{\large \it
${}^{(a)}$ Department of Physics, Ben-Gurion University, Beer-Sheva
84105,
Israel\\
${}^{(b)}$ Theory Division, CERN, CH-1211, Geneva 23, Switzerland\\
${}^{(c)}$ Department of Physics, University of Pennsylvania, \\
\vspace*{0.5ex}
Philadelphia, PA 19104, U. S. A.}

\vspace*{4.5ex}
{\bf Abstract}
\end{center}
\vspace*{3.0ex}

We report on progress towards evaluation of stringy
non-perturbative effects, using a two dimensional effective
field theory for matrix models. We briefly discuss the relevance of
such effects to models of dynamical supersymmetry breaking.

\vspace{2.8cm}
\noindent
\rule[.1in]{16.5cm}{.002in}\\
\noindent
$^{*)}$ Contribution to the proceedings, based on a talk given at
International Workshop on Supersymmetry and Unification of Fundamental
Interactions (SUSY 95), Palaiseau, France, 15-19 May 1995.
\\

\vfill
\begin{flushleft}
CERN-TH/95-289 \\
October 1995 \end{flushleft}

\newpage
\
\setcounter{page}{1}
\pagestyle{plain}

\section{Introduction}

Supersymmetry breaking, particularly in the framework of string theory,
is an
interesting unsolved theoretical problem whose solution may have  observable
low-energy consequences accessible to future laboratory experiments. It is
widely believed that dynamical supersymmetry breaking in string theory occurs
through non-perturbatively induced interactions. At present it is
possible to
evaluate and control non-perturbative interactions  in the low-energy
field theory approximation to string theory which are typically of
strength $exp(-1/g^2_{string})$, where $g_{string}$ is the string coupling
constant. These interactions are an essential ingredient in  models of
supersymmetry breaking in
string theory, as described, for example, in  Taylor's talk.

However, it has been known for a while \cite{shenker}, that in addition
to
non-perturbative interactions of strength $exp(-1/g^2_{string})$ there
are also
the so called ``stringy non-perturbative effects" of strength
$exp(-1/g_{string})$. Recently, some proposals for the source of stringy
non-perturbative effects were put forward, one proposal \cite{mgjp} is
that the
their source are certain ``$D$-instantons", associated with
disconnected world-sheet holes and another proposed source \cite{witten}
are  type II string solitons of mass $1/g_{string}$.

Our proposal \cite{rbbo} is that stringy non-perturbative effects  are
associated
with classical solutions for which the string coupling parameter varies in
space-time and becomes strong in some region. In the space-time region of
strong
coupling, new degrees of freedom are important and are the source of stringy
non-perturbative effects. Our method of calculation of these effects is
based on
a two dimensional effective field theory approach, and therefore can be
applied
in cases in which the effective dynamics in the strong coupling region is
two-dimensional.

At the moment, the relationship between the different proposed sources  is
unclear, but the  possibility that some or all have
a common origin is very interesting. We will have nothing to add on this
subject in this talk, rather we describe in some detail our approach.

\section{Effective Two Dimensional Theory}

Collective field theory \cite{das} for bosonic $d=1$ matrix models
\cite{done} is written in terms of
the density of matrix eigenvalues, $\der_x\vphi=\sum\limits_{i=1}^{N}
\d(x-\l_i(t))$, where  $\l_i(t)$ are the matrix eigenvalues, and the
dimension
of the matrix $N$, is very large. The (Euclidean) action  of collective field
theory is written in terms of the collective field $\vphi$, and is given by
\bq S_E[\vphi]=\int
dx dt \bl\frac{\dot{\vphi}^2}{2\vphi'}+\frac{\pi^2}{6}\vphi^{'3}
    +\hf (\frac{1}{\w g}-\w^2 x^2)\vphi'\br.
 \label{collact}\eq
The static, high density solution of the equation of motion derived from the
action (\ref{collact}) is denoted by $\phi_0\equiv\der_x\vphi_0$, and is
given by the simple expression  $\phi_0=\w/\pi\sqrt{x^2-1/\w g}$ for the
range
$|x|\ge\sqrt{1/\w g}$. There are also interesting time dependent Euclidean
solutions in the low density region $|x|\le\sqrt{1/\w g}$, which will be
discussed later.

Action (\ref{collact}) has two notable deficiencies. First, the kinetic
term is not in canonical form, signifying that the correct canonical
field of
the theory is not $\vphi$, and second, the coordinate $x$ appears
explicitly in
the potential and therefore Poincare invariance seems to be broken
explicitly.
We remove both deficiencies. The first, following ref. \cite{das}, by
expanding
around the classical solution, $\der_x\vphi=\phi_0+\der_x\z/\sqrt{\pi}$ and
changing coordinates $x\rightarrow \tau=\int\limits^x dy/(\pi\phi_0)$.
The resulting action for $\z$ is given by
\bq S_E[\z] = \int dt d\tau\bl
       \hf(\dot{\z}^2+\z^{'2})
       -\hf\frac{\gbig(\tau)\dot{\z}^2\z'}{1+\gbig(\tau)\z'}
       +\frac{1}{6}\gbig(\tau)\z^{'3}
       -\frac{1}{3}\frac{1}{\gbig(\tau)^2}\br,
 \label{zest} \eq
where $\gbig(\tau)$ is a space dependent coupling parameter, which we
define below, and the $\tau$ integration is over the limits
$-\infty<\tau\le \tau_0-\frac{\s}{2}$
and $\tau_0+\frac{\s}{2}\le\tau<\infty$. In $\tau$ space,
the low density  region is given by
$\tau_0-\frac{\s}{2}<\tau<\tau_0+\frac{\s}{2}$, so that
$\tau_0$ is the center of the low density region and $\s$ is the
width.
The coupling parameter, defined over
$-\infty<\tau\le \tau_0-\frac{\s}{2}$ and
$\tau_0+\frac{\s}{2}\le\tau<\infty$,
is given by $\gbig(\tau)=(\pi^{3/2}{\vphi}_0(x))^{-1}$, and
is found to be
\bq \w \gbig(\tau)=
     4\sqrt{\pi} g \frac{ e^{-2\w(\tau-\tau_0-\frac{\s}{2})}}
     {(1-e^{-2\w(\tau-\tau_0-\frac{\s}{2})})^2},
 \label{ggdef}\eq
for the range $\tau_0+\frac{\s}{2}\le\tau<\infty$, with the obvious symmetric
form in the range $-\infty<\tau\le \tau_0-\frac{\s}{2}$.  Notice that the
coupling parameter blows up as $\tau\rightarrow\tau_0\pm\frac{\s}{2}$;
that is,
at the boundaries of the low density region.

We turn now to correct the second deficiency of action (\ref{collact}) and
restore Poincare invariance. We interpret, following
ref.\cite{ramshanta}, the
space dependent coupling parameter $\gbig(\tau)$ as a field dependent
coupling
parameter $\gbig(D)$. We further assume that the field $D$ has a space
dependent expectation value $\langle D\rangle$, such that
$\gbig(\tau)=\gbig(\langle D\rangle)$. Furthermore, we impute the apparent
lack of Poincare invariance of the action (\ref{zest}) solely to the space
dependence of the expectation value $\langle D\rangle$.  It turns out that,
although not unique, a Poincare invariant action $S_E[\z,D]$  does exist
and is
not arbitrary. In its simplest form it is given by
\begin{eqnarray}
S_E[\z,D]\!\!\!\! &=&\!\!\!\! \int d^2 X \Biggl\{
\hf(\nabla \zeta)^2 +\frac{1}{4}
{(\nabla \zeta)^2\nabla \zeta\!\!\cdot\!\!\nabla D\over 1-1/(2\w^2)
\gbig(D){\nabla\zeta\!\!\cdot\!\!\nabla D}}
-\frac{1}{48 \w^4} \gbig(D)
(\nabla \zeta\!\!\cdot\!\!\nabla D)^3   \nonumber\\ & + &
\frac{1}{32 \w^4} \gbig(D)
{(\nabla \zeta\!\!\cdot\!\!\nabla D)^3\over 1-1/(2\w^2)
\gbig(D){\nabla\zeta\!\!\cdot\!\!\nabla D}}-
\frac{1}{24}\frac{1}{\gbig^2(D)}
\left[ {(\nabla D)^2+4\w^2} \right]
\Biggr\} \label{efflag}
\end{eqnarray}
where the coupling $\gbig(D)$ is given by
\bq  \gbig(D)=
     4\sqrt{\pi} g \frac{ e^{D}}
     {(1-e^{D})^2}.
\label{gddef}\eq
Thus $g$ cannot be absorbed into a redefinition of $D$ and scaled away
completely from the effective action, although asymptotically, for
$D\ll-1$ it
is possible to scale it away.

Action (\ref{efflag}) may look complicated, but it actually encodes
simple and interesting dynamics. For $D\ll -1$ the coupling parameter
$\gbig(D)$ is negligibly small and our system reduces to two decoupled
fields, one massless field $\z$, and one superheavy field $D$. Note also
that $\z$ has only derivative interactions and therefore no potential. As
the value of $D$ increases the interaction strength increases
exponentially until the coupling blows up for $D=0$, signaling the
spontaneous generation of a boundary. The width of the strongly coupled
region (the ``wall") is $1/\w$. The conclusion is therefore that the
action (\ref{efflag}) describes, effectively, a free massless scalar
field $\z$, moving in a bounded region of space-time.

The equations of motion derived from (\ref{efflag}) are quite complicated
but a complete and simple set of classical solutions may be found in a
straightforward manner,
\bq
D_0= A X_1 + B X_2 +C, \hspace*{2cm} \z_0=c
\label{dsol}
\eq
where $A,B,C,c$ are real parameters and $A^2+B^2=4 \w^2$.
Each $D_0$ solution in (\ref{dsol}) defines a line along which the
coupling parameter blows up and a boundary is formed. The solutions we
are interested in are a combination of two
solutions for which the two boundaries are parallel to each other and at
a constant fixed width.  It is convenient to define the directions
parallel and perpendicular to $D_0$,
$\widehat X _{\parallel}=\frac{1}{2\w}(AX_1+BX_2)$ and
$\widehat X _{\perp}=\frac{1}{2\w}(-B X_1+AX_2)$. We observe that
Poincare invariance is not completely broken by the solutions
(\ref{dsol}). One translation in the  ${X_\perp}$ direction remains
unbroken.
The energy $E_0$ of the classical solutions, which can be computed using
standard methods, is found to vanish. In fact, the whole energy-momentum
tensor $T_{\mu\nu}$ vanishes. Note that to compute correctly $T^{\mu\nu}$
one has to correctly couple gravity to the system, as done for example in
\cite{ramshanta}, compute $T^{\mu\nu}$ in a curved background and then
set space-time to be flat. Reduction of action (\ref{efflag}) to
collective field theory action is achieved by setting $D$ to one of its
possible expectation values $\langle D\rangle=D_0$ and identifying
$X_\parallel$ with $\tau$ and $X_{\perp}$ with $t$. The effective action
(\ref{efflag}) reduces exactly to the corresponding action (\ref{zest})
for each classical solution.

Have we successfully restored Poincare invariance to our theory? It
certainly seems so, but we have to be careful! Because the action
(\ref{efflag}) is Poincare invariant and the $D_0$ solutions break some
of that invariance there should be zero-modes corresponding to the broken
generators of Poincare (Euclidean) invariance. In our case they are a
rotation, and a translation in the $X_{\parallel}$ direction. Indeed one
finds that the expected zero-modes exist.
For example, the wave function corresponding to the broken generator
$\der_{X_\parallel}$ is simply proportional to $\der_{X_\parallel}
D_0=-2\w$. The standard argument about symmetry restoration then says
that each classical solution does break some of the symmetry, but the
symmetry is restored by the summation over zero-modes. However, the
standard argument holds only  if all the zero-modes are normalizable. Of
course, the correct measure, determined by the kinetic terms, has to be
used to compute the normalization factor. When we compute the
normalization factor $N$, for the ${X_\parallel}$ translation zero-mode,
for example, we find that it is divergent $N^2\sim \int dX_{\parallel}
1/\gbig^2(X_\parallel)$. When a non-normalizable zero-mode appears the
theory breaks up into separate superselection sectors parametrized by the
value the broken generator takes in each sector. In particular this would
mean that Poincare invariance was not really fully restored in our theory
(see, for example, a discussion of For example, the wave function
corresponding to the broken
Note that the divergence of $N$ comes from the weak coupling region, and
therefore to achieve actual Poincare invariance restoration we need to
regulate the weak coupling behavior of $D$ in some way. At the moment we
cannot offer a conclusive opinion on whether and how the weak coupling
behavior of $D$ can be modified and regulated. We can look for clues by
understanding  how a higher-dimensional string theory handles a similar
challenge. Let us look, for example, at the 5-dimensional extremal black
hole solution of the heterotic string \cite{caetal}
\begin{equation}
ds^2= -Q dt^2+(1+{Q\over r^2}) (dr^2 +r^2 d\Omega^2_3), \hspace*{1cm}
e^{2(D-D_0)}=  1+{Q\over r^2}
\label{clsol}
\end{equation}
where $H= Q \epsilon_3$ is the solution for field strength of the
antisymetric tensor $B$ and  $D$ is the dilaton.
In the ``throat" region, ${Q\over r^2}>>1$ the solution is approximately
\begin{equation}
ds^2\sim -Q dt^2 +Q d\tau^2 +Q d\Omega^2_3  \hspace*{1cm}
D-D_0\sim  - \tau
\end{equation}
where $\tau\sim \ln r$.
Many other similar examples in different dimensions can be found in the
review \cite{dkl}. It is important to note that the dilaton  (the analog
of our field $D$) in the exact solution is asymptotically constant. It
varies linearly only in the ``throat" region.  The translation zero-modes
corresponding to the broken translation generators in the background of
the classical solution (\ref{clsol}) are normalizable since the would-be
weak coupling divergence is regulated by the constant non-zero
asymptotic value of the dilaton.
Consider the expansion
of the heterotic string effective action around the classical solution in
the ``throat" region \cite{giddstro}.
Because the geometry of this region is that of $M_2\times S_3$, the light
fields can be described by an effective two dimensional field theory in
$(t,\tau)$
space. This theory is of course not Poincare invariant because the
dilaton has a space dependent expectation value. The coupling parameter of
the theory varies in space for the same reason.
The light fields of the 2-d theory are just the modes of the antisymmetric
tensor which in this case can be described by one derivatively
coupled scalar field, the axion.
The similarity between our 2-d theory and the one associated with regions
of linear dilaton solutions of the heterotic string  suggests a physical
way to regulate the weak coupling divergence by modifying the behavior of
$D$ from linear to constant asymptotically.

\section{Stringy Non-perturbative effects}

As mentioned previously, in addition to the static, high density,
solution of the collective field theory equations of motion discussed in
the previous section, there are interesting time-dependent Euclidean
solutions in the low density region. In the effective field theory the
low density region is formally a region of infinite coupling and the
important degrees of freedom are not smooth excitations of the fields,
but rather single matrix eigenvalues, corresponding to singular field
configurations (see also \cite{dmw}). To expose the important physics in
the low density region we separate one discrete eigenvalue from the
continuum and look at it's effective dynamics,
\bq L_E[\l_0;\vphi] = \hf\dot{\l}_0^2+
\hf\w^2(\frac{1}{\w g}-\l_0^2)
     +\int dx\frac{\vphi'}{(x-\l_0)^2}
     + \int dx\bl\frac{\dot{\vphi}^2}{2\vphi'}
     +\frac{\pi^2}{6}\vphi^{'3}
     +\hf\w^2(\frac{1}{\w g}-x^2)\vphi'\br.
 \label{hybrid}\eq
The third term in this expression represents the mutual
interaction of the discrete eigenvalue with the continuum eigenvalues,
which are collectively described using the classical solution $\phi_0$.
We obtain the Euclidean equations of motion for $\l_0$ by variation of
(\ref{hybrid}), they are given
in the small $g$ limit simply  by
\bq \ddot{\l}_0+\w^2\l_0 = 0;\  -1/\sqrt{\w g}<\l_0<1/\sqrt{\w g}
\hspace*{1cm}\ddot{\l}_0 = 0;\  \l_0=\pm 1/\sqrt{\w g}.
 \label{keq} \eq
We also impose the following boundary conditions,
$\l_0(t\rightarrow -\infty)= \pm 1/\sqrt{\w g}$ and, independently,
$\l_0(t\rightarrow +\infty)= \pm 1/\sqrt{\w g}$.
There are two static solutions to (\ref{keq}) which satisfy this
boundary condition,
\bq {\l}_0^{(\pm)}=\pm 1/\sqrt{\w g}\sin\w(t-t_1)
    \hspace{.2in} ; \hspace{.1in}
    t_1-\frac{\pi}{2\w}\le t\le t_1+\frac{\pi}{2\w},
 \label{klink}\eq
representing tunneling of single eigenvalues across the potential barrier
in the low density region.

The partition function associated with the theory discussed above can be
written as a sum over different instanton sectors and
after some lengthy  analysis \cite{bfo}, using a dilute gas
approximation, we arrive at
the following general
result
  \brr Z &=& \int[d\vphi]e^{-S_\vphi[\vphi]}\sum_{q=0}^\infty
     \frac{1}{q!}\M^q\prod_{i=1}^q
     \int dt_i\sum_{\{k_i\}}\prod_{j=1}^q
     e^{-S_I^{(k_j)}[\vphi;t_j]} \nonumber \\
     &=& \int[d\vphi]e^{-S_\vphi[\vphi]}\sum_{q=0}^\infty
     \frac{1}{q!}
     \bl\M\int dt_1\bpl e^{-S_I^{(+)}[\vphi;t_1]}
     +e^{-S_I^{(-)}[\vphi;t_1]}\bpr\br^q.
 \err
The sum over $q$ is now an exponential, so that
$Z=\int[d\vphi]e^{-S_{eff}[\vphi]}$,
where $S_{eff}[\vphi]=S_\vphi[\vphi]+\Delta S[\vphi]$ is the effective
action with the instanton effects systematically
incorporated, and
\bq \Delta S[\vphi]=\M\int dt_1
    \bl e^{-S_I^{(+)}[\vphi;t_1]}
     +e^{-S_I^{(-)}[\vphi;t_1]}\br
 \label{delta}\eq
is the associated change in the action.
The action  $S_I^{(\pm)}$ is given by
\bq
S_I^{(\pm)}[\vphi;t_j]=
\int_{t_j-\frac{\pi}{2\w}}^{t_j+\frac{\pi}{2\w}}dt
\int dx\bl\frac{\vphi'(x,t)}{(x-\l_0^{(\pm)}(t-t_j))^2}
-\frac{\vphi'(x,t)}{(x-\l_{\so}^{(\pm)}(t-t_j))^2}\br.
 \label{dali}\eq
where
\bq \l_{\so}^{(\pm)}(t;t_1)=\left\{\begin{array}{ll}
      \mp 1/\sqrt{\w g} & ;\hspace{.2in} t_1-\frac{\pi}{2\w}\le t<t_1 \\
      \pm 1/\sqrt{\w g} &  ;\hspace{.2in} t_1<t\le t_1+\frac{\pi}{2\w}
\end{array}\right..
 \label{step}\eq
The quantity $\M$ is a dimensionful parameter that sets the
basic strength for induced non-perturbative interactions
\bq \M=\w\sqrt{\frac{\pi}{2g}} e^{-\frac{\pi}{2g}}.
\label{madef}\eq

A full analysis to find the induced operators in collective field theory
action was carried out \cite{bfo}. We give only the final result for the
simplest induced operator.
\bq \D S[\z]=2\w g^{-1/6}e^{-\frac{\pi}{2g}}\int dt
     e^{-\frac{2\sqrt{2}}{3\w}\z'(\tau_0,t)}.
 \eq
Note that the the induced operator contains only a $t$ integration
which appears  because of the existence of a normalizable $t$ translation
zero-mode. Similarly, once a good regularization procedure is found to
render the two other zero-modes normalizable, the induced operators due
to our instantons would be integrated against a Poincare-invariant
integration measure and therefore themselves be Poincare invariant
completions of the computed operators we discussed above.
\section{Concluding remarks and Outlook}

We described a method for the evaluation of stringy non-perturbative
effects and their systematic inclusion in the form of induced operators
into an effective action. We have not yet completed the calculation due
to a missing regularization procedure which has to be developed. When we
successfully find a regularization scheme we may turn to the evaluation
of similar effects in the supersymmetric effective theory presented in
\cite{bfos} and also in higher dimensional theories, if these theories
contain strong coupling regions that are effectively two dimensional, as
in the example presented in section 2. It would be interesting to find
out in which form the operators in the supersymmetric theory appear. One
possibility is that stringy non-perturbative effects induce
supersymmetric operators, which may affect the pattern of supersymmetry
breaking, the other possibility is that the induced operators  break
supersymmetry. Preliminary indications suggest that the latter
possibility, but final conclusions have to We described a method for
\section{Acknowledgment}
Research  supported in part by the Department of Energy under  contract
No. DOE-AC02-76-ERO-3071.

 \end{document}